\documentclass[runningheads,orivec]{llncs}
\usepackage[T1]{fontenc}
\usepackage{hyperref}
\usepackage[table,xcdraw]{xcolor}

\usepackage{rotating}
\usepackage{amssymb}
\usepackage{scalerel}
\usepackage[ruled]{algorithm2e}
\usepackage{bbding}
\usepackage{booktabs}
\usepackage{amsmath}
\usepackage[shortlabels]{enumitem}
\usepackage{AdditionalCommands}

\definecolor{CVDdarkred}{HTML}{c1272d}
\definecolor{CVDindigo}{HTML}{0000a7}
\definecolor{CVDteal}{HTML}{008176}

\DeclareRobustCommand\orcidlink[1]{%
  \texorpdfstring{\href{https://orcid.org/#1}{{\raisebox{-.5ex}{\includegraphics[height=1.1em]{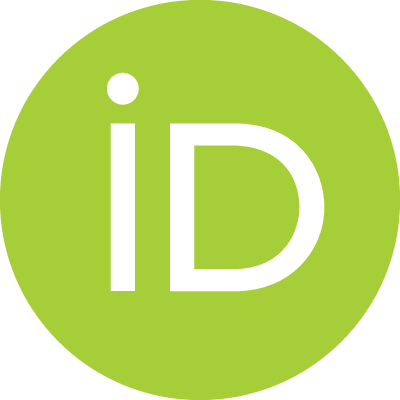}}}}}{https://orcid.org/#1}%
}

\begin{document}

\title{\texorpdfstring{\estSq}{eST2} Miner - Process Discovery Based on Firing Partial Orders}

\author{Sabine Folz-Weinstein\inst{1}\href{mailto:sabine.folz-weinstein@fernuni-hagen.de}{\Envelope}\orcidlink{0009-0004-6946-4011}
\and 
Christian Rennert\inst{2}\orcidlink{0000-0003-4614-6171}
\and
Lisa Luise Mannel\inst{2}\orcidlink{0000-0001-6158-356X}
\and
\mbox{Robin Bergenthum\inst{3}\orcidlink{0000-0003-0464-8843}}
\and
Wil van der Aalst\inst{2}\orcidlink{0000-0002-0955-6940}
}

\authorrunning{S. Folz-Weinstein et al.}

\institute{Chair of Data Science, University of Hagen, Germany
\email{sabine.folz-weinstein@fernuni-hagen.de}\and
Chair of Process and Data Science (PADS), RWTH Aachen University, Germany
\email{\{rennert,mannel,wvdaalst\}@pads.rwth-aachen.de} \and
Faculty of Mathematics and Computer Science, University of Hagen, Germany
\email{robin.bergenthum@fernuni-hagen.de}}

\maketitle

\begin{abstract}
Process discovery generates process models from event logs.
Traditionally, an event log is defined as a multiset of traces, where each trace is a sequence of events. 
The total order of the events in a sequential trace is typically based on their temporal occurrence. However, real-life processes are partially ordered by nature. Different activities can occur in different parts of the process and, thus, independently of each other. Therefore, the temporal total order of events does not necessarily reflect their causal order, as also causally unrelated events may be ordered in time. 
Only partial orders allow to express concurrency, duration, overlap, and uncertainty of events. Consequently, there is a growing need for process mining algorithms that can directly handle partially ordered input. 
In this paper, we combine two well-established and efficient algorithms, the eST Miner from the process mining community and the Firing LPO algorithm from the Petri net community, to introduce the \estSq Miner. The \estSq Miner is a process discovery algorithm that can directly handle partially ordered input, gives strong formal guarantees, offers good runtime and excellent space complexity, and can, thus, be used in real-life applications.

\end{abstract}

\section{Introduction}
\begin{figure}[t]
    \centering
    \includegraphics[width=0.9\linewidth]{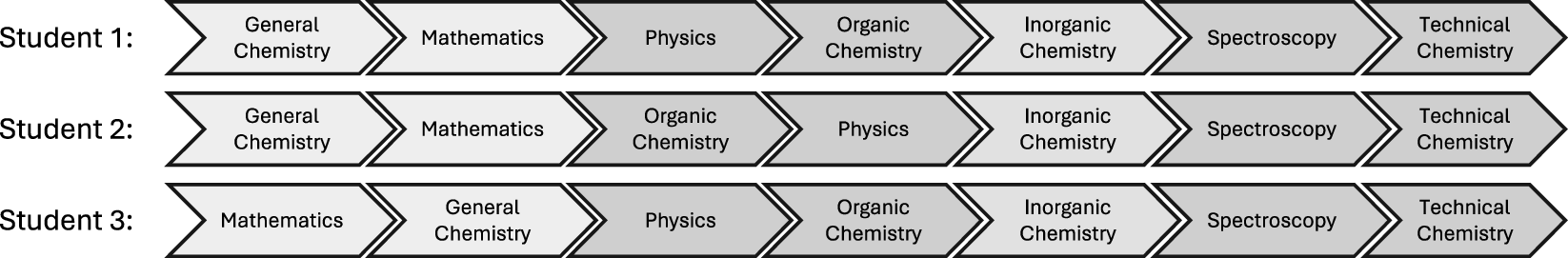}
    \caption{Example event log with three traces. Each trace is a sequence of courses taken by one student, totally ordered based on course exam dates.}
    \label{fig:edutotal}
\end{figure}

Process mining gains insights into business processes by analyzing recorded behavior \cite{task_force_manifesto}. 
The goal of process discovery is to generate a process model based on an event log
\cite{aalst_PM_discovery_conformance_enhancement_BP,aalst_Data_Science_in_Action,aalst_Process_Mining_Handbook}.
An event log is a multiset of traces, where each trace is a sequence of events, i.e., executed activities of a process. Traditionally, these traces are totally ordered based on the timestamps of the events.
To illustrate this, consider an example from an Educational Process Mining project at the RWTH Aachen which analyzes study behavior \cite{bogarin_educationalpm,rennert_studyplans}.
Here, every trace in the event log represents the sequence of courses that a student took, ordered by the timestamps when the student passed the course exams.
Figure \ref{fig:edutotal} shows three traces of the event log, where
student 1 took \textit{General Chemistry}, \textit{Mathematics}, etc. in the depicted order. Note that these traces suggest that, e.g., \textit{Technical Chemistry} is always completed after \textit{Spectroscopy} because this relation exists in all traces.

\begin{figure}[t]
    \centering
    \includegraphics[width=0.5\linewidth]{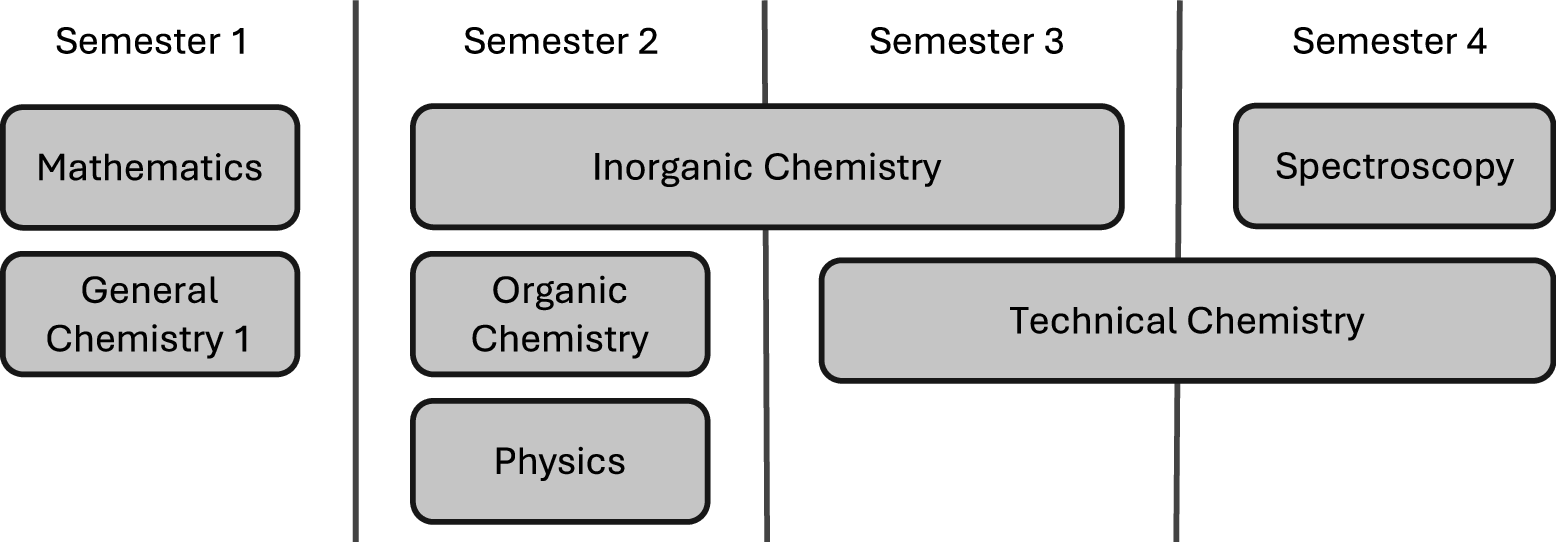}
    \caption{Partially ordered representation of the three traces in Figure \ref{fig:edutotal}.}
    \label{fig:edupartial}
\end{figure}

In real-life processes, however, activities are often only partially ordered, and some activities can occur independently of each other. 
Consequently, an observed temporal order of the events does not necessarily reflect their causal relation, as also causally unrelated events may be ordered in time. 
In our example, we know that students usually do not take courses in a strict sequential order (one after the completion of another), but take several courses in parallel per semester, and that courses have a duration. Thus, if we group all exam timestamps by semester and include course durations, we receive the partially ordered trace in Figure~\ref{fig:edupartial} for all traces of Figure~\ref{fig:edutotal}, depicting that the students took several courses concurrently.
Note that the order relation between \textit{Spectroscopy} and \textit{Technical Chemistry}, inferred by the totally ordered traces, is accidental and caused by later exam dates within the same semester.
The information that \textit{Technical Chemistry} overlaps with \textit{Inorganic Chemistry} and, thus, is not dependent on the completion of \textit{Inorganic Chemistry} could not be reflected by the totally ordered traces either.

This example illustrates that if we discover models based on event data that is totally ordered based on timestamps, on the one hand, we may unwillingly infer dependencies between activities to the discovered model which are purely accidental.
On the other hand, we may lose valuable information on the underlying process that is available in the event log, but a total order cannot represent.
The more concurrent behavior there is in the underlying process, the higher the risk that the discovered model will not depict the process correctly.

Quite often, the temporal order based on the timestamps of the events is not a total order either. Most real-life event logs have severe data quality problems, leading to timestamps that are unreliable, incomparable, have too coarse or different granularities, especially if data from different source systems and/or manual input must be combined \cite{aalst_wanna_improve,terhofstede_true_frontier_of_pm}. A well-known example is healthcare data \cite{mans_processmining_in_healthcare,Martin_dataquality_in_pm}, where many events require manual input to the system, usually done at the end of a shift. Consequently, the timestamp does not necessarily reflect the occurrence of the event itself. Moreover, several events can have the same timestamp, e.g., a date only. Many BPI challenge event logs illustrate the same problem: In the BPIC 2011 log, 87\%, and in the BPIC 2012 log, 5\% of all events have the same timestamp as their predecessors \cite{aalst_may_i_take_your_order}. In this case, the events are partially ordered due to uncertainty, and any total order would be random.

In turn, there is an increasing amount of additional data available in information systems which can be used to identify causal relations within a process. An example is lifecycle data, which reflects the duration and overlap of events (e.g., contained in the BPIC 2012 log). A lifecycle attribute for events is defined in the event log standard format XES \cite{DBLP:journals/cim/WynnAVS24}, but this information cannot be represented using totally ordered traces.

Thus, to obtain meaningful process mining results and discover models that better reflect the underlying process, we find a growing amount of work in which a trace is a partial order of events~\cite{aalst_may_i_take_your_order,bergenthum_prime_miner,dumas_process_mining_reloaded_event_structures_unified_representation,folz-weinstein_ILP2,lu_partial_order_survey,carmona_haar_unfolding_based_discovery,pegoraro_discovery}.
Using partial orders, we can explicitly express both uncertainty and concurrency~\cite{koutny_structure_concurrency,pratt15modelling}, as well as represent duration and time overlaps of events.
Furthermore, partial orders often provide a much more compact representation of the recorded behavior, which can improve runtime and space complexity of discovery algorithms.

From a practical point of view, the use of partially ordered event logs has three consequences concerning process discovery applications.
First, we need to distinguish between data in event logs that reflects technical information, i.e., generated or required by the information system, and data that reflects characteristics and dependencies of the underlying process.
Second, we need an additional preprocessing step for partial order extraction/event log transformation, using data identified in the first step.
Third, we need process discovery algorithms that can directly process partially ordered input in real-life settings, which is the focus of this paper. Applying sequential trace-based algorithms on partially ordered event logs is not an efficient option, because we must process all possible interleavings (i.e., sequentializations) of all partially ordered traces, and one partially ordered trace can induce a significant number of interleavings. Furthermore, semantically, concurrency is not the same as interleaving.

In~\cite{lu_partial_order_survey}, the authors present an overview of partial order-based process discovery.
The existing work primarily refers to synthesis or folding, e.g., Prime Miner~\cite{bergenthum_prime_miner}, \ilpSq Miner~\cite{folz-weinstein_ILP2}, unfolding-based process discovery~\cite{carmona_haar_unfolding_based_discovery}, multi-phase process mining approaches~\cite{dongen_multi_phase_process_mining}, and folding-based approaches~\cite{bergenthum_folding_partially_ordered_runs}.
These approaches give strong formal guarantees, i.e., they produce models with high fitness and precision. However, they come with considerable space and runtime complexity, which is problematic when working with real-life event logs.

To close this gap, in this paper, we combine two well-established and efficient algorithms, the eST Miner and the \firing algorithm from Petri net theory, to introduce the \estSq Miner. The eST Miner \cite{mannel_basic_est_miner_complex_token_game} is a replay-based process discovery algorithm. 
To find the places of the result Petri net, it enumerates and evaluates all possible places of the net in linear time by firing every trace in the event log. Thanks to a special strategy for ordering, traversing, and pruning the set of possible candidate places, the algorithm is time-efficient and only needs to store the input event log and the resulting net.
Working with partially ordered event logs, however, it is not possible any more to simply replay traces from start to end. Therefore, in the \estSq Miner, we adapt the currently most efficient verification algorithm from Petri net theory~\cite{bergenthum_firing_partial_orders} to verify whether a partial order is replayable.
The new \estSq Miner handles totally ordered event logs just like the eST Miner and provides the same guarantees for the discovered process models, but it can also directly handle partially ordered input.

In this paper, we address the problem of discovering a process model based on an event log which is a multiset of labeled partial orders. We introduce the \estSq Miner which is based on two well-established algorithms. We implement the \estSq Miner and evaluate the new approach based on public and private logs. 
\section{Preliminaries}
A multiset $m$ over $\setX$ is a function $m \colon \setX \rightarrow \naturals$. We write $m = \sum_{\elemx\in\setX} m(\elemx) \cdot \elemx$ to denote all multiplicities of $m$.
We extend the notion of a subset to the concept of multisets, where a multiset is considered a subset of another multiset if the cardinality of every element is equal to or less than that in the other multiset. 

We model observed partially ordered behavior as a partially ordered set of activities (see \cite{aalst_Process_Mining_Handbook} for a detailed introduction).
\begin{definition}[Labeled Partial Order, lpo]
    Let \setA be a set of activities.
    A~labeled partial order (lpo) is a triple $(\partialNodes, \partialRelation, \labelingfunc)$ where \partialNodes is a finite set of nodes, ${\partialRelation \subseteq \partialNodes \times \partialNodes}$ is a transitive and irreflexive relation, and $\labelingfunc \colon \partialNodes \rightarrow \setA$ a labeling function.
    Let $\elemn \in \partialNodes$ be a node.
    We denote $\aset{\elemn' \in \partialNodes \mid \elemn' \partialRelation \elemn}$ the set of predecessors and $\aset{\elemn' \in \partialNodes \mid \elemn \partialRelation \elemn'}$ the set of successors of \elemn.
\end{definition}

We define an event log as a multiset of labeled partial orders.

\begin{definition}[Event Log]
    An event log is a multiset of labeled partial orders.
\end{definition}

To model business processes, we use the concept of workflow nets \cite{aalst_workflow}, a subclass of marked Petri nets \cite{desel_reisig_place_or_transition_petri_nets}.

\begin{definition}[Workflow Net]
    A workflow net \net is a tuple $(\places, \transitions, \arcs)$ where \places is a finite set of places, \transitions is a finite set of transitions such that $\places \cap \transitions = \emptyset$ holds, $\arcs\subseteq (\places \times \transitions) \cup (\transitions \times \places)$ is a set of directed arcs, $i,o \in \places$ are two places for which $\preset{i} = \postset{o} = \emptyset$ holds, and where every node $n \in (\places \cup \transitions)$ is on a directed path from $i$ to $o$. Let $\elemn \in (\transitions \cup \places)$ be a node of a workflow net.
    We denote ${\preset{\elemn} = \aset{\elemn' \in (\transitions \cup \places) \mid (\elemn', \elemn) \in \arcs}}$ the preset of $\elemn$ and $\postset{\elemn} = \aset{\elemn' \in (\transitions \cup \places) \mid (\elemn, \elemn') \in \arcs}$ the postset of \elemn.
\end{definition}

For workflow nets, there is a simple firing rule.
A marking of \net is a multiset $\marking \colon \places \rightarrow \naturals$.
A transition \transition can fire at marking \marking if $\preset{\transition} \subseteq \marking$ holds.
Once transition~\transition fires, the marking of \net changes from \marking to $\marking'$,  where for $\marking'$ it holds that $\marking'(p) = \marking(p) - 1$ if $\place \in (\preset{\transition} \setminus \postset{\transition})$ or  $\marking'(p) = \marking(p) - 1$ if $\place \in (\postset{\transition} \setminus \preset{\transition})$  or $\marking'(p) = \marking(p)$ otherwise.

The behavior of a workflow net is the set of all possible partially ordered sets of firing events that bring the workflow net from its initial marking \textit{i} to its final marking \textit{o}. Formally we can define this language as the set of labeled partial orders for which a so-called valid tokenflow exists for every place \cite{bergenthum_firing_partial_orders}. A \textit{compact} tokenflow is a distribution of tokens on the skeleton arcs of the lpo. This distribution is valid if it satisfies certain conditions. In the following, we adopt the original definition of compact tokenflows to workflow nets.

\newlength{\originaltextfloatsep}
\setlength{\originaltextfloatsep}{\textfloatsep}
\setlength{\textfloatsep}{-10mm}

\makeatletter
\tagsleft@true
\makeatother
\makeatletter
\renewcommand{\tagform@}[1]{\maketag@@@{\textit{(#1)}}}
\makeatother
\renewcommand{\theequation}{\roman{equation}}

\begin{definition}[Behavior of a Workflow Net]\label{def:compact_tokenflow}
    Let $\net = (\places, \transitions, \arcs)$ be a workflow net, let $lpo = (\partialNodes,\partialRelation, \labelingfunc)$ be a labeled partial order, and $\labelingfunc(\partialNodes) \subseteq \transitions$. Let < be the smallest relation for which the transitive closure is \partialRelation. A compact tokenflow is a function $\funcTokenFlow \colon < \rightarrow \naturals$. Fix a $\place \in  \places \setminus \{i,o\}$.
    Place p is valid for lpo if and only if there is a compact tokenflow x such that the following conditions hold:
    \begin{enumerate}[label={(\textit{\roman*})}]
        \item $\begin{aligned}[t]
            \partialNode \in \partialNodes, (\place,\labelingfunc(\partialNode)) \in \arcs \implies \sum_{\partialNode' < \partialNode} \funcTokenFlow(\partialNode', \partialNode) \geq 1 \text{, and}
            \label{def:compact_tokenflow_i}
        \end{aligned}$
        \item $\begin{aligned}[t]
        \forall_{\partialNode \in \partialNodes} \colon 
        \sum_{\partialNode \partialRelation \partialNode'} \funcTokenFlow(\partialNode, \partialNode') =
            \begin{cases}
              \sum_{\partialNode' \partialRelation \partialNode} \funcTokenFlow(\partialNode', \partialNode) - 1,  (\place,\labelingfunc(\partialNode)) \in \arcs \land (\labelingfunc(\partialNode),p) \notin \arcs,\\
             \sum_{\partialNode' \partialRelation \partialNode} \funcTokenFlow(\partialNode', \partialNode) + 1,   (\labelingfunc(\partialNode),p) \in \arcs \land (p,\labelingfunc(\partialNode)) \notin \arcs,\\
             \sum_{\partialNode' \partialRelation \partialNode} \funcTokenFlow(\partialNode', \partialNode)\hspace{\widthof{+ 1}},\text{ otherwise}.
            \end{cases}
        \end{aligned}$
        \label{def:compact_tokenflow_ii}
    \end{enumerate}
    Place i is valid for lpo if and only if
    \begin{enumerate}[label={(\textit{\roman*})},resume]
        \item $\begin{aligned}[t]
        \big| \{v \in V \mid (i,l(v)) \in F \} \big| = 1.
        \label{def:compact_tokenflow_iii}
        \end{aligned}$
    \end{enumerate}
    Place o is valid for lpo if and only if
    \begin{enumerate}[label={(\textit{\roman*})},resume]
        \item $\begin{aligned}[t]
        \big| \{v \in V \mid (l(v),o) \in F \} \big| = 1.
        \label{def:compact_tokenflow_iv}
        \end{aligned}$
    \end{enumerate}
If and only if all places of \net are valid for lpo, lpo is in the language of \net.
\end{definition}

\setlength{\textfloatsep}{\originaltextfloatsep}

\renewcommand{\theequation}{\arabic{equation}}
\makeatletter
\tagsleft@false
\makeatother

Condition \ref{def:compact_tokenflow_i}~ensures that every transition receives enough tokens to fire, \ref{def:compact_tokenflow_ii}~that the firing rule applies, \ref{def:compact_tokenflow_iii}~and \ref{def:compact_tokenflow_iv}~that the initial token is consumed and the final token is produced. 

An event log is replayable in a workflow net if every lpo of the event log is in the language of the workflow net.

\begin{figure}[t]
    \centering
    \includegraphics[width=0.7\linewidth]{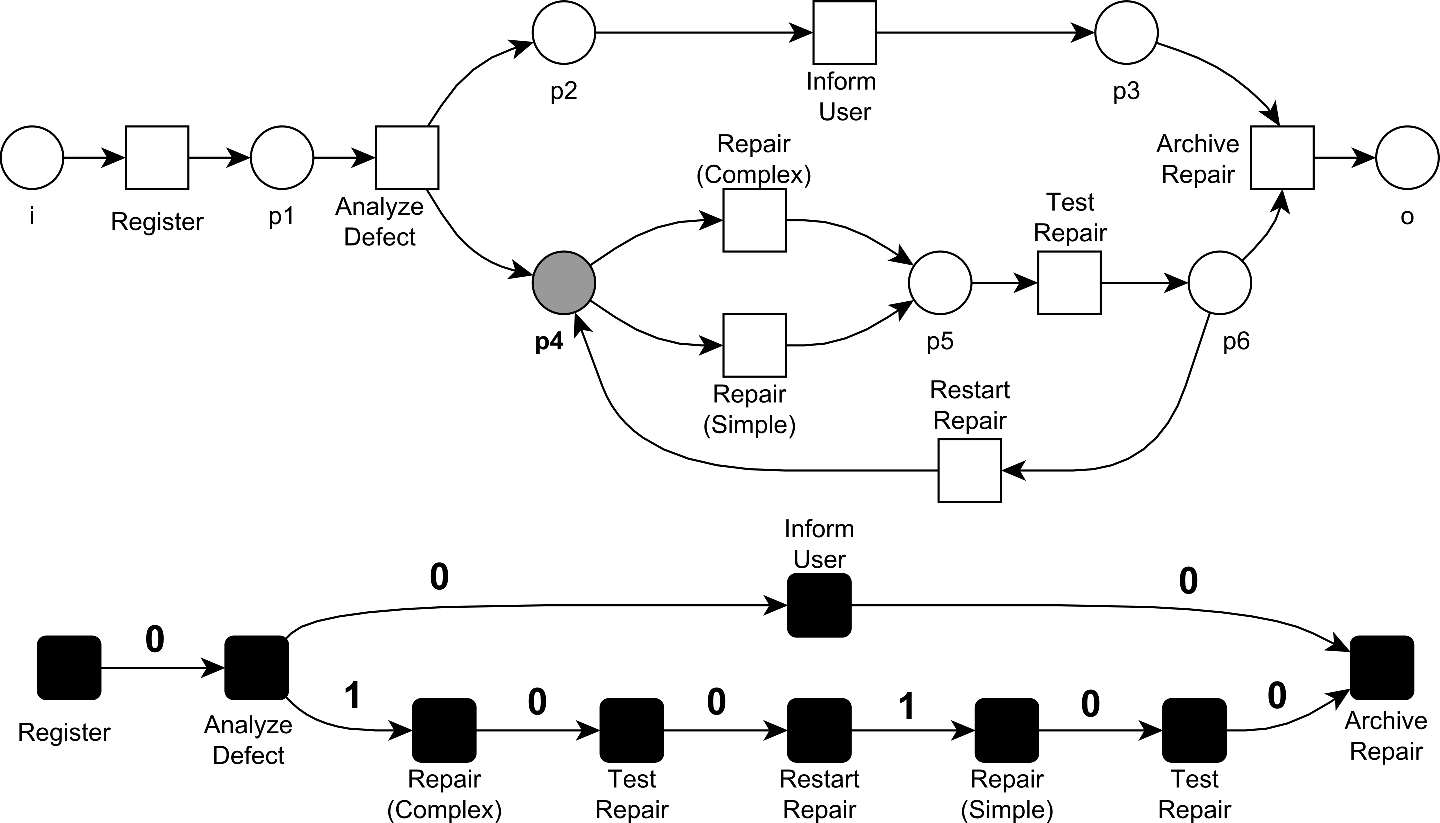}
    \caption{A workflow net (top) and a labeled partial order with a compact tokenflow for p4 (bottom).}
    \label{fig:validct}
\end{figure}

\textbf{Example 1.}
Figure \ref{fig:validct} shows a workflow net and an lpo.
The numbers denoted on the arcs of the lpo depict a compact tokenflow for place \textit{p4}.
In this tokenflow, both \textit{Repair (Complex)} and \textit{Repair (Simple)}, which are in the postset of \textit{p4}, receive enough tokens to fire (Def. \ref{def:compact_tokenflow}\ref{def:compact_tokenflow_i}).
The firing rule (Def. \ref{def:compact_tokenflow}\ref{def:compact_tokenflow_ii}) applies because \textit{Analyze Defect} and \textit{Restart Repair}, which are in the preset of \textit{p4}, increase the amount of tokens that they received from their predecessors in the lpo by one token.
\textit{Repair (Complex)} and \textit{Repair (Simple)}, which are in the postset of \textit{p4}, decrease the amount by one token.
All other events leave the amount unchanged.
Def. \ref{def:compact_tokenflow}\ref{def:compact_tokenflow_iii} and Def. \ref{def:compact_tokenflow}\ref{def:compact_tokenflow_iv} are satisfied because in \net, there are arcs from \textit{i} to \textit{Register} and from \textit{Archive Repair} to \textit{o}, such that the initial token can be consumed and the final token be produced.
Thus, all conditions are fulfilled and \textit{p4} is valid for this lpo. 
\section{Replay-Based Process Discovery}
In this section, we recapitulate the basic ideas of replay-based process discovery and the original eST Miner as presented in \cite{mannel_basic_est_miner_complex_token_game}, which serves as the foundation for our new \estSq Miner approach.

To start with, we initialize a workflow net by simply creating one transition for each activity in the event log.
This initial placeless net can replay any given event log because no place restricts the firing of the transitions.
Then, we successively add places and related arcs to prune the behavior of the workflow net such that it matches the behavior of the given event log.
This idea is similar to discovery algorithms based on the theory of regions \cite{aalst_Process_Mining_Handbook}.

To find the places of the workflow net, we enumerate and traverse all possible places of the net. In this context, we assume that each place also defines its preset and postset, i.e., how it connects to a set of transitions.
The set of all possible places of the net is called the set of candidate places. For each candidate place, we evaluate if it is valid for the given event log by replaying the event log on the place.
Only if the place is valid, it can be added to the final result.
If a candidate is not valid, we move on to check the next candidate place.

The main idea of the eST Miner is that, during this evaluation of a candidate place, we do not only evaluate if a place is valid or not.
For places which are not valid, we distinguish if this is due to a lack of tokens or due to remaining tokens on the place.
Thus, while replaying each trace in the event log on the candidate, we keep track of the number of tokens in every intermediate marking.
If the number of tokens ever becomes negative, the behavior is not replayable, and we call the candidate place \textit{underfed}.
If, after replaying the behavior of the event log, the place is not empty, we call the candidate place \textit{overfed}.
Note that a place can be underfed and overfed at the same time.
The additional underfed and overfed information is used for an efficient traversal of the set of candidates. 

Although the set of potential candidate places is finite, it is still exponential in the number of transitions, a number impossible to efficiently store, retrieve, and evaluate.
Therefore, the eST Miner deterministically calculates the next candidate place to be evaluated and prunes the candidate space based on the evaluation results of the current candidate place.
To achieve this, the set of candidate places is represented in the form of a tree, with different branches reflecting the addition of incoming and outgoing arcs to transitions.
If a candidate place is underfed, adding outgoing arcs will not help make it a valid candidate place.
Similarly, if a candidate is overfed, adding incoming arcs will not help make it valid.
Using the underfed and overfed information of the current candidate place, the algorithm can prune the tree and skip the traversal of entire unfitting subtrees.
This drastically decreases the overall runtime on real-life event logs (by $40-95 \%$), still guaranteeing that all valid places are visited. 

By applying such a replay-based process discovery strategy, we only add valid places to the net.
As a consequence, the language of the discovered workflow net will always include the behavior of the event log, and the model will be perfectly fit.
However, for practical applications, this is too restrictive, as event logs are known to also contain noise. To address this, the eST Miner uses an additional, second evaluation step on log level and a configurable noise handling parameter~$\tau$.
In the log level evaluation step, the results of all traces of the event log are summarized.
Only if a place is valid for a fraction of $\tau$ or more of the traces in the event log, the place is added to the result net.
The log level candidate evaluation using $\tau$ is also applied on the underfed and overfed values used for the pruning and traversal of the candidate tree. 

\begin{figure}[t]
    \centering
    \includegraphics[width=0.85\linewidth]{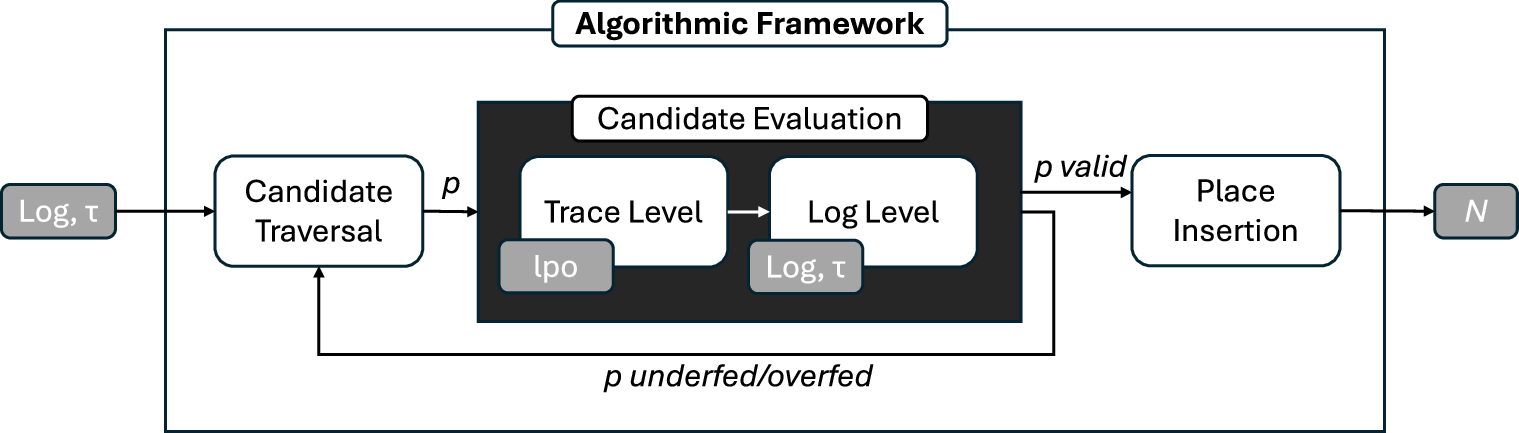}
    \caption{High-level overview of the \estSq Miner framework.}
    \label{fig:eSTOverview}
\end{figure}

A high-level overview of the algorithmic framework is provided in Figure~\ref{fig:eSTOverview}. 
For implementation details of the original eST Miner, including post-processing of the result net, we refer the interested reader to \cite{mannel_basic_est_miner_complex_token_game,mannel_implicit_places}. 
\section{The \estSq Miner}
To apply the concept of replay-based process discovery to partially ordered event logs, we introduce the \estSq Miner.
The \estSq Miner follows all the concepts outlined in the previous section.
Thus, the \estSq Miner is a variant of the eST Miner, but it addresses the candidate evaluation specifically for labeled partial orders.
It decides whether a candidate place is valid, underfed, or overfed for a given lpo by calculating compact tokenflows (Def. \ref{def:compact_tokenflow}).
Since we evaluate each lpo of the entire partially ordered event log on every candidate place, the runtime and efficiency of this evaluation part are vital.

To verify whether a valid compact tokenflow exists for a place and a given lpo, a maximal flow problem must be solved in $\mathcal{O}(n^3)$ time \cite{bergenthum_firing_partial_orders}, where $n$ is the number of nodes of the lpo.
The maximal flow algorithm explores all possible distributions of tokens on the arcs of the lpo, and directly indicates if a candidate place is valid, underfed, or overfed.
However, solving a maximal flow problem in cubic time is not an efficient option for large real-life event logs.

Thus, for the \estSq Miner, we adapt the currently most efficient approach presented in \cite{bergenthum_firing_partial_orders}, which we will refer to as the \textit{\firing} algorithm in the remainder of this paper. 
The main idea of the \firing algorithm is that for certain places, it is easy to determine whether a valid compact tokenflow exists for a given lpo or not.
To identify such places, the algorithm applies two heuristics steps first, the so-called \textit{forward} and \textit{backward strategies}.
Both of these heuristics steps run in linear time.
Consequently, the exact maximal flow algorithm only needs to be applied on places which cannot be resolved by the heuristics steps. 

Experimental results of the \firing algorithm show that it runs in quadratic or even linear time for most practical use cases \cite{bergenthum_firing_partial_orders}, and that it is especially fast on restricted Petri nets like workflow nets.
In workflow nets, there are no arc weights, and a lot of places are empty most of the time.
Therefore, the number of possible distributions of tokens decreases significantly, and the probability of finding a valid compact tokenflow in the two heuristics steps increases considerably.
Therefore, in most cases, the replayability of an lpo on a place can be decided after forward or backward strategy, i.e., in linear time (like the token replay strategy for totally ordered traces in the original eST Miner).

Note that it is also possible to evaluate whether totally ordered traces are replayable using the \firing algorithm.
In totally ordered traces, only one single possible distribution of tokens exists.
Therefore we can always decide whether a place is valid or not after executing the first heuristics step, which in fact is equivalent to the original eST Miner replay strategy~\cite{mannel_basic_est_miner_complex_token_game}.
For detailed information on the \firing algorithm, we refer the interested reader to \cite{bergenthum_firing_partial_orders}.

\textbf{Necessary Adaptations.}
Although certain aspects of the original \firing algorithm must be adapted to the scope and setting of the \estSq Miner, the two algorithms are quite a perfect match.

The \textit{input} to our adapted \firing algorithm within the \estSq Miner is an lpo and a workflow net which only consists of the candidate place \textit{p} and the transitions in its preset and postset.
As the original \firing algorithm was designed for general marked Petri nets, and workflow nets are a restricted form of general marked Petri nets, these one-place nets can be handled by the algorithm without any modifications. 

However, it is no longer sufficient to evaluate whether a place is valid or not for a specific lpo.
Therefore, we must adapt the \textit{output} of our adapted algorithm to provide the more specific information whether the place is underfed and/or overfed.
This means, if a place is not valid, we need to distinguish if this is due to missing or remaining tokens on the place. 

The original algorithm verifies if a place is valid or not for a specific lpo by evaluating whether or not a compact tokenflow (i.e., a distribution of tokens along the arcs of the lpo) exists such that every node of the lpo receives enough tokens for its related transition to fire.
If no such distribution exists, one or several nodes of the lpo do not receive enough tokens, which can be directly translated to the place being underfed.

The original algorithm does not evaluate yet if tokens remain on a place, i.e., whether or not a place is overfed, because this is irrelevant for general marked Petri nets. 
However, as a coincidence, the original Firing LPO algorithm already calculates the final marking, i.e., the amount of tokens which are not consumed by the last node of the lpo, and which remain on the place.
This final marking is proven to be unique even if no valid compact tokenflow can be constructed \cite{bergenthum_firing_partial_orders}.
We can use this final marking to identify if a place is overfed.

\textbf{Place Evaluation Using the adapted Firing LPO algorithm.}
The \estSq Miner extends every lpo of the partially ordered event log by a unique start node~\sactivity, which is earlier than all other nodes of the lpo, and a unique final node~\eactivity, which is later than all other nodes of the lpo.
In our result workflow net, we connect the unique place~\textit{i} to \sactivity and the unique place~\textit{o} to \eactivity such that $\postset{i} = \aset{\sactivity}$ and $\preset{o} = \aset{\eactivity}$ hold.
Thus, we ensure that the initial token on~\textit{i} is consumed and the final token on~\textit{o} is produced (Def. \ref{def:compact_tokenflow}\ref{def:compact_tokenflow_iii} and \ref{def:compact_tokenflow_iv}) by construction, and that we only evaluate inner places of a workflow net. 

The evaluation starts with the \emph{forward strategy} heuristics.
In this heuristics step, we process all nodes of the lpo in one sequential order which respects the $\partialRelation$-order.
We brute-force construct one possible distribution of tokens along the arcs of the lpo, i.e., one possible compact tokenflow.
The algorithm verifies if every node receives enough tokens to fire (Def. \ref{def:compact_tokenflow}\ref{def:compact_tokenflow_i}), and ensures that the firing rule applies (Def. \ref{def:compact_tokenflow}\ref{def:compact_tokenflow_ii}).
To guarantee a linear runtime, the algorithm can only explore one possible distribution of tokens in the heuristics step, i.e., one compact tokenflow, but it identifies and stores the information if alternative distributions are possible.
At the end of the forward heuristics step, it calculates the unique final marking. 

After the forward strategy heuristics step, we can always decide if a place is \textbf{overfed} or not. A place is overfed if the final marking is positive; otherwise, it is not overfed. 
For some places, we can also decide if they are \textbf{underfed} or not:

\begin{enumerate}
    \item If the final marking is negative, we can deduce that in all possible distributions of tokens, there will be a lack of tokens.
    Thus, it is impossible to fulfill Def. \ref{def:compact_tokenflow}\ref{def:compact_tokenflow_i}, and we can classify the place as underfed.
    \item If we found a distribution such that for each node of the lpo, there are enough tokens for its related transition to fire, then Def. \ref{def:compact_tokenflow}\ref{def:compact_tokenflow_i} is fulfilled for all nodes of the lpo, and the place is not underfed.
    \item If Def.\ref{def:compact_tokenflow}\ref{def:compact_tokenflow_i} is violated for one or several nodes of the lpo, we must still consider whether the algorithm detected that alternative distributions of tokens are possible.
    If no alternative distributions are possible, the place is underfed. 
\end{enumerate}
We cannot decide whether the place is underfed in case Def. \ref{def:compact_tokenflow}\ref{def:compact_tokenflow_i} is violated, but alternative distributions are possible.
In this case, one of these alternative distributions may be a valid distribution.
Therefore, we apply the second heuristics step of the algorithm, the \textit{backward strategy}.
The backward strategy works just like the forward strategy but in the reverse direction, i.e., it processes all nodes of the lpo in the reverse total order used in the forward strategy, except that it does not calculate and evaluate a final marking anymore.

If no decision can be made by the backwards heuristics step either, the maximal flow algorithm with worst-case cubic runtime must be applied to decide whether or not the place is underfed.
\section{Evaluation}

\begin{sidewaysfigure}
    \centering
    \includegraphics[height=9.8cm]{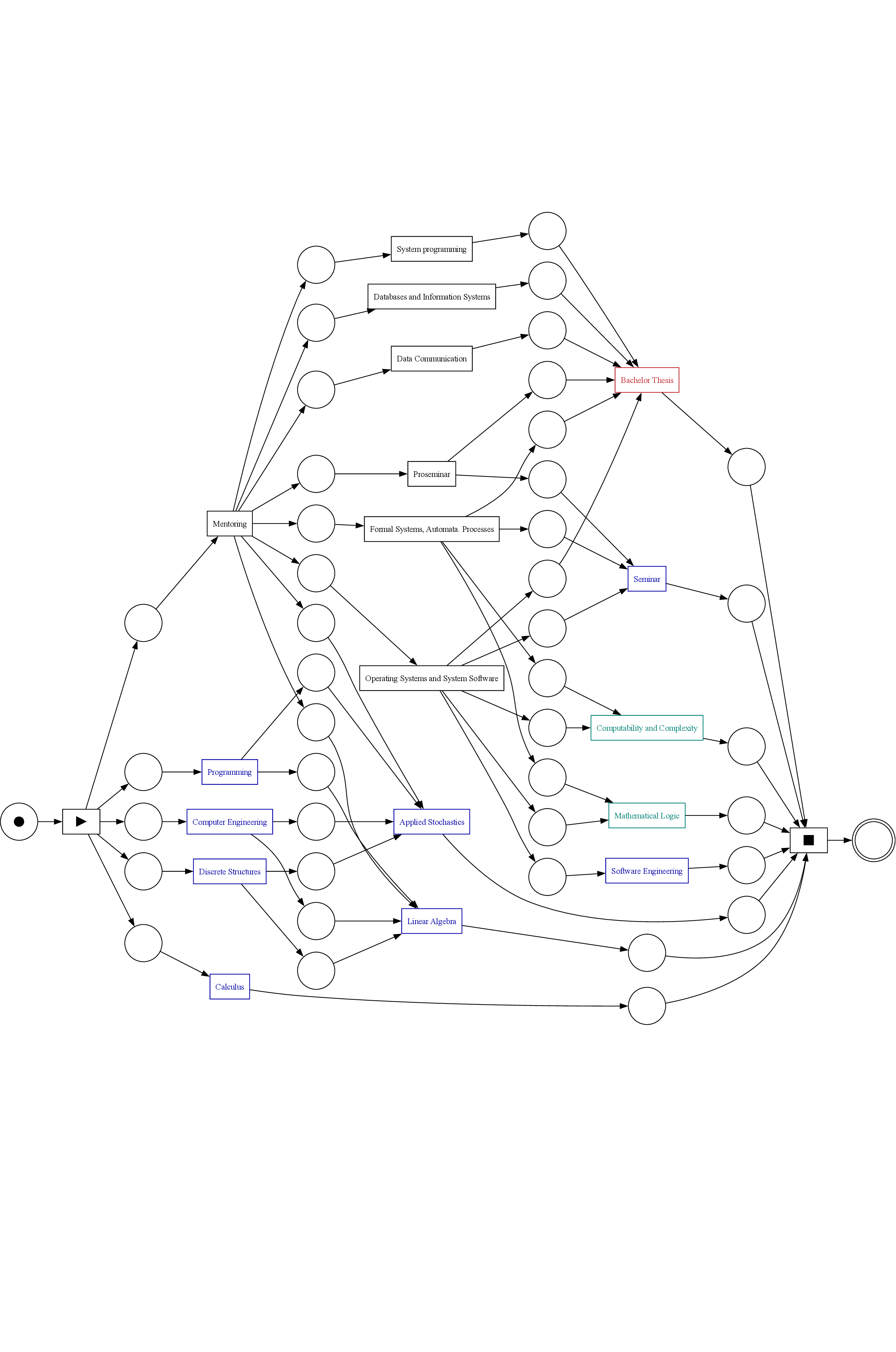}\hspace{1.5cm}
    \includegraphics[height=9.8cm]{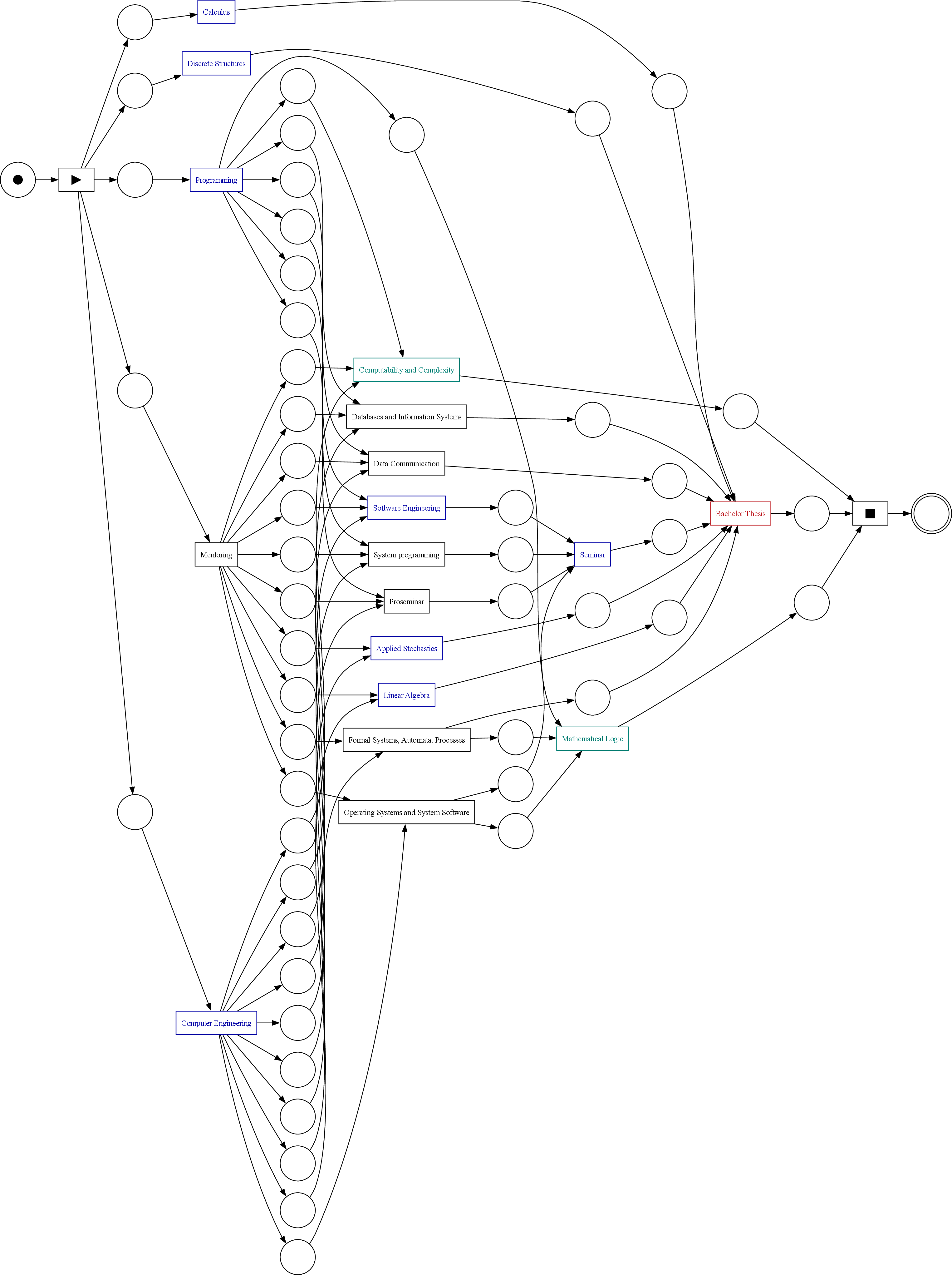}
    \caption{Workflow nets discovered by the \estSq Miner based on a partially ordered event log (left) and by the eST Miner based on a totally ordered event log (right) containing the study behavior data of 355 students of the RWTH Aachen which completed their Bachelor's degree of Computer Science and started their studies in winter semesters 2015-2018.}
    \label{fig:edu-nets}
\end{sidewaysfigure}

Resuming our Educational Process Mining example, Figure \ref{fig:edu-nets} depicts two workflow nets discovered by the \estSq Miner on a partially ordered (left) and by the eST Miner on a totally ordered event log (right) based on study behavior data.
The depicted nets are not intended to be readable in detail, but to highlight their structural differences.
As expected, significantly more dependencies exist in the net that was discovered based on totally ordered input (right).
As an example, consider the \textit{Bachelor~Thesis} marked in \textcolor{CVDdarkred}{red}.
In the net based on totally ordered input (right), we find order relations between the \textit{Bachelor Thesis} and eight other courses, marked in \textcolor{CVDindigo}{blue}, while only two courses, marked in \textcolor{CVDteal}{teal}, are concurrent to \textit{Bachelor~Thesis}.
In the net discovered on partially ordered input (left), all of these courses (marked \textcolor{CVDteal}{teal} and \textcolor{CVDindigo}{blue}) are concurrent to \textit{Bachelor Thesis}.
Project owners confirm that students frequently complete courses while already working on their bachelor thesis.
Presumably, the order relations are found because the bachelor colloquiums are usually held at the very end of a semester, after all other course exams.
Since the event log data of this project is not public and the project is still ongoing, we are unable to further evaluate the quality of these two process models in more detail.
As the project progresses, a more detailed validation will certainly provide more valuable insights.

Therefore, to evaluate our approach in a transparent, reproducible way and on a broader basis, we use several well-known public totally ordered event logs which we transform to partially ordered event logs using a so-called concurrency oracle.
Concurrency oracles~\cite{dumas_local_concurrency_detection_in_BP_event_logs,bergenthum_prime_miner,lu_partial_order_survey} use heuristics to derive a partial order relation on top of the total order relation, based on additional attributes of the event log.
As the focus of our evaluation is not on the performance of oracles, we apply the same oracle on all our test event logs. 
We use the so-called Alpha~oracle \cite{bergenthum_prime_miner} implemented in the CCO tool\footnote{\url{https://github.com/sabinefw/ConfigurableConcurrencyOracleTool}} \cite{folz2024partially}, which evaluates directly-follows relations to identify concurrency.
Note that deriving partially ordered logs from a totally ordered event log using a concurrency oracle may infer some additional behavior that was not recorded in the event log, such that process models discovered based on the two log versions are not entirely comparable.
However, due to the characteristics of the Alpha oracle, the partially ordered version of the log essentially contains the same behavior as the totally ordered event log but in another representation.
Each partial order represents several traces of the totally ordered log (i.e., sequentializations of the partial order).
Thus, we expect to discover the same process models for the two event log versions. 

To this date, no comparable process discovery approaches for partially ordered input exist that run on real-life event logs, nor do conformance checking metrics for partially ordered input.
Thus, to evaluate our approach, we compare the \estSq Miner (based on the partially ordered version of our test event logs) to the eST Miner (based on the totally ordered version).
This comparison still allows us to assess the overall quality of the discovered process models discovered by the \estSq Miner, as well as to compare the runtimes.
We evaluate the quality of the workflow nets discovered by the \estSq Miner using standard quality metrics based on the behavior described by the totally ordered event log version, since no other metrics exist yet.
Note that we compare the runtimes of \estSq Miner and eST Miner mainly to assess the general efficiency of the \estSq Miner approach.
As the main goal of partial-order based process discovery is not a speedup, but the ability to directly process partially ordered input for the reasons described in the introduction, we are interested in a comparison of the algorithm runtimes in relation to the sizes of the event log versions.
Often, partial orders are a more compact representation of the behavior, since several total orders may be interleavings of the same partial order.

\textbf{Implementation.}
The \estSq Miner is implemented in ProM\footnote{\url{https://promtools.org/}}, built on the basis of the most recent implementation of the eST Miner \cite{DBLP:journals/fuin/MannelA24}, accessible on GitHub\footnote{\url{https://github.com/promworkbench/eST2-miner}}.
Note that specific parameters required by this implementation remain fixed during all our experiments ($\delta = 1$, $s = 5$, candidate space tree depth of~$5$).
To exclude a possible runtime impact with respect to different existing eST Miner implementations, we use the \estSq Miner implementation on all event log versions (totally and partially ordered), since for totally ordered traces, the trace evaluation step in the \estSq Miner is equivalent to the replay in the eST Miner.

\textbf{Experimental Setup.}
The evaluation is performed single-threaded on a 16-core Intel i7-1260P 2.10 GHz machine with 32 GB of main memory. 

\begin{table}[t]
    \centering
    \caption{Fitness and precision values of the workflow nets discovered by the \estSq~Miner with $\tau$ thresholds of 1.0, 0.8, and 0.5.}
    \label{tab:metrics}
    \begin{tabular}{l|cc|cc|cc} 
        \toprule
                   & \multicolumn{2}{c|}{$\tau = 1.0$} & \multicolumn{2}{c|}{$\tau = 0.8$} & \multicolumn{2}{c}{$\tau = 0.5$} \\ 
                   event log & fitness    & precision   & fitness  & precision     & fitness & precision     \\ 
        \midrule
        Repair     & 1.00    & 0.64         & 1.00     & 0.64           & 0.88       & 0.84            \\
        Teleclaims & 1.00    & 0.31        & 0.96     & 0.53            & 0.82       & 0.40           \\
        Reviewing  & 1.00    & 0.48        & 1.00     & 0.48            & 0.97       & 0.49           \\
        RTFM       & 1.00    & 0.15        & 0.94     & 0.68           & 0.79       & 0.51      \\
        BPI12(a)   & 1.00    & 0.20        & 0.95     & 0.35           & 0.80       & 0.79            \\
        BPI12(o)   & 1.00    & 0.20       & 0.97     & 0.24           & 0.85       & 0.87             \\
        BPI19(c)   & 1.00    & 0.14        & 0.94     & 0.44           & 0.88       & 1.00            \\
        Sepsis(40) & 1.00    & 0.09        & 0.99     & 0.14           & 0.97       & 0.16         \\
        \bottomrule
        \hline
    \end{tabular}
\end{table}
\begin{table}[t]
    \centering
    \caption{Runtime in seconds for the \estSq Miner on the partially ordered event log version and for the eST Miner on the totally ordered event log version with $\tau = 1.0$.}
    \label{tab:runtime}
    \begin{tabular}{l|r|rr|r|rr|r} 
        \toprule
        \multicolumn{1}{c|}{}& \multicolumn{1}{c|}{} & \multicolumn{2}{c|}{\#variants} &\multicolumn{1}{c|}{relative}&  \multicolumn{2}{c|}{runtime in s}     \\ 
        event log            &    \#cases               & lpos                 & traces  & difference & \estSq & eST   & speedup             \\ 
        \midrule
        Repair            & 1,000                       & 9                    & 39 & 77\% & 1.03     & 4.59       & 78\%                        \\
        Teleclaims        & 3,512                       & 8                    & 12 & 33\%&  2.91     & 4.71       & 38\%                 \\
        Reviewing         & 100                         & 93                   & 96 & 3\%& 204.38   & 202.35     &  -1\%              \\
        RTFM              & 150,370                     & 85                   & 231 & 63\%& 25.00    & 83.05     &  70\%             \\
        BPI12(a)          & 13,087                      & 12                   & 17 & 29\%& 1.96     & 3.44       &   43\%               \\
        BPI12(o)          & 5,015                       & 75                   & 168 & 55\%& 7.48     & 15.54       &  52\%             \\
        BPI19(c)          & 14,498                      & 206                  & 281 & 27\%& 135.37   & 172.61      &   22\%           \\
        Sepsis(40)        & 1,050                       & 435                  & 846  & 49\%& 1,074.61 & 1,768.46   &  39\%          \\
        \bottomrule
    \end{tabular}
\end{table}

We use the artificial event logs Repair example\footnote{Example event log file for: \url{https://promtools.org/}.} and Teleclaims~\cite{aalst_Data_Science_in_Action} as well as the real-life event logs Reviewing~\cite{aalst_Data_Science_in_Action}, road traffic fine management (RTFM), Sepsis, and the BPI Challenge logs 2012 (A), 2012 (O) and 2019\footnote{Online accessible at: \url{https://data.4tu.nl/}.} for our experiments.
The BPI Challenge log 2019 is filtered for the ``Consignment'' trace attribute, and the Sepsis log for traces up to a maximal length of 40. 

\textbf{Fitness and Precision.}
Table \ref{tab:metrics} depicts the fitness and precision values for the workflow nets discovered by the \estSq Miner with noise handling thresholds $\tau$ of $1.0$, $0.8$, and $0.5$.
This means that (at least) $100\%$, $80\%$, or $50\%$ of the cases in the event log must be replayable in the result net.
We used fitness metrics based on alignments, and precision metrics based on escaping edges \cite{aalst_Process_Mining_Handbook}.
The fitness of the result nets is related to the selected $\tau$ thresholds, and the precision tends to decrease with increasing fitness.
Both values can be balanced by selecting a suitable $\tau$.
Note that for almost all test event logs, using the same parameter settings, the same workflow nets are discovered for both event log versions.
In case the discovered nets differ, the difference is minimal. 

\textbf{Runtime.}
Table \ref{tab:runtime} compares the runtime in seconds for \estSq~Miner and eST~Miner for a fixed $\tau$ threshold of $1.0$.
As a reference, we include the number of cases, lpo and sequential trace variants for each event log, to assess the size of the partially and the totally ordered input. 
For most event logs, the speedup of the \estSq~Miner compared to the runtime of the eST Miner scales almost linearly to the relative difference between the log versions. 
For example, in the Teleclaims log, the relative difference between the traces in the log variants is 33\%, almost linearly reflected in the speedup of 38\%. 
Even in the Reviewing event log, where there are almost as many lpos as totally ordered traces, the \estSq~Miner does not show a significant runtime increase compared to the eST~Miner.
This supports our expectation that the candidate evaluation step of the \estSq Miner is executed in linear time for most candidate places, like in the eST Miner (although the problem is now cubic), and that the \estSq Miner approach is comparably efficient.
\section{Conclusion}
There are various important reasons for using partially ordered event logs. Real-life processes are partially ordered by nature, and there are ways to obtain data on the causal relation of the events.
A total order based on the temporal order is prone to fail as soon as the timestamps are unreliable, incomparable, or too coarse granular. 
Only partial orders allow us to directly model uncertainty, concurrency, duration, and overlap of events, which a total order cannot capture. 
Overcoming the total order assumption is especially relevant in fields such as healthcare, education, and logistics which exhibit highly concurrent behavior. Consequently, there is a growing need for process mining algorithms which can directly handle partially ordered input. 

To bridge this gap and fully exploit the concurrency information of real-life data in process discovery, we combine two well-established and efficient algorithms, the eST Miner process discovery algorithm and the \firing algorithm from Petri net theory, to introduce the \estSq Miner. The \estSq Miner is a process discovery algorithm which can handle both partially and totally ordered input while maintaining the same guarantees with respect to the discovered workflow nets as the established eST Miner, offering space efficiency and good runtimes even on real-life event logs.
We conducted several experiments with well-known public event logs, assessing the runtime of the algorithm and the quality of the discovered process models. The results show that, for the majority of our experiments, the runtime of the \estSq Miner on partially ordered input scales almost linearly with the level of compactification of the event log compared to the eST Miner, and that the same nets are discovered based on both event log versions. 

In future work, we plan to evaluate the \estSq Miner with respect to other new partial order-based discovery approaches, all based on the same partially ordered input, to further analyze, compare, and optimize our approach.
We also intend to conduct further experiments to evaluate and analyze the quality of the discovered process models based on partially ordered event logs created from the same totally ordered event log using different concurrency oracle types.

\begin{credits}
    \subsubsection{\ackname}
    We thank the Alexander von Humboldt (AvH) Stiftung for supporting our research (grant no. 1191945).
    The authors gratefully acknowledge the financial support by the Federal Ministry of Education and Research (BMBF) for the joint projects AIStudyBuddy (grant no. 16DHBKI016) and Bridging AI (grant no. 16DHBKI023).
    The authors have no competing interests to declare that are relevant to the content of this article.
\end{credits}

\bibliographystyle{splncs04}
\bibliography{mybibliography}

\end{document}